# Strongly Absorbing Nanoscale Infrared Domains within Graphene Bubbles

Tom Vincent,[†,‡] Matthew Hamer,[§,∥] Irina Grigorieva,[§,∥] Vladimir Antonov,[¶,‡] Alexander Tzalenchuk[†,‡] and Olga Kazakova*[,†]

[†]National Physical Laboratory, Hampton Road, Teddington TW11 0LW, U.K.
[‡]Department of Physics, Royal Holloway University of London, Egham TW20 0EX, U.K.
[§]School of Physics and Astronomy, University of Manchester, Manchester M13 9PL, U.K.
[∥]National Graphene Institute, University of Manchester, Manchester M13 9PL, U.K.
[¶]Skolkovo Institute of Science and Technology, Nobel str. 3, Moscow, 143026, Russia

**ABSTRACT:** Graphene has shown great potential for modulating infrared (IR) light in devices as small as 350 nm. At these length scales, nanoscale features of devices, and their interaction with light, can be expected to play a significant role in device performance. Bubbles in van der Waals heterostructures are one such feature, which have recently attracted considerable attention thanks to their ability to modify the optoelectronic properties of 2D materials through strain. Here we use scattering-type scanning near-field optical microscopy (sSNOM) to measure the nanoscale IR response from a network of variously shaped bubbles in hexagonal boron nitride (hBN)-encapsulated graphene. We show that within individual bubbles there are distinct domains with strongly enhanced IR absorption. We correlate this with strain in the graphene, found with confocal Raman microscopy and vector decomposition analysis. This reveals intricate and varied strain configurations, in which bubbles of different shape induce more bi- or uniaxial strain configurations. Ridges in the bubbles, seen by atomic force microscopy (AFM), coincide with the domain boundaries, which leads us to attribute the domains to nanoscale strain differences in the graphene. This reveals pathways towards future strain-based graphene IR devices.

Graphene and hexagonal boron nitride (hBN)'s unique optoelectronic properties make them ideally suited for a variety of applications in the infrared (IR) range.[1,2] In graphene, surface plasmon polaritons (SPPs; coupled oscillations of light and free charge carriers) have been shown to have wavelengths many orders of magnitude smaller than the diffraction limit, long lifetimes and exceptional electronic tunability via gating.[3–10] Similarly in hBN, hyperbolic phonon polaritons (HPhPs; coupled oscillations of light and optical phonons) have short wavelengths and long lifetimes, as well as negative phase velocities.[11–14] When the two materials are layered together in van der Waals heterostructures they support hybrid phonon-plasmon polaritons,[15] as well as even longer polariton lifetimes[16] and moiré modulated polariton dispersions.[17] These properties have been utilised in a range of optoelectronic devices, including photodetectors operating in the IR[18] and THz[19] regions, and optical modulators with footprints as small as 350 nm.[20,21]

Recently, a great deal of attention has been paid to the effects of strain on 2D materials. Compared to bulk materials, their electronic and optical properties are especially susceptible to modification by strain. This, combined with their significant pliability, has led to a new field of research dubbed *straintronics*.[22] Examples in the infrared regime include demonstrations that the dispersion of HPhPs is altered by the presence of strain in hBN,[23] and that wrinkles in graphene may act as scattering sites for SPPs.[24]

Bubbles in van der Waals heterostructures have emerged as an interesting platform to study the effects of strains on 2D materials.[25] In transition metal dichalcogenides, bubbles have been shown to act as highly localised photoluminescent emitters, with strain-dependent peak energies.[26] In graphene, nanoscale bubbles have been shown to act as localised plasmonic hotspots,[27] and to sustain high-Tesla pseudomagnetic fields.[28]

These bubbles are formed due to competition between van der Waals and elastic potential energies, in the presence of interlayer contamination, which may be formed of adsorbed hydrocarbons and water vapour.[29] This effectively squeezes the contamination into pockets, leaving micron-scale areas with atomically sharp interlayer interfaces. This process, referred to as *self-cleaning*, promotes interlayer adhesion enabling large-area van der Waals heterostructures to be realised.[29,30]

A benefit of using bubbles to strain 2D materials is that the stochastic nature of their formation means they provide a wide range of strain values and configurations. This means that, with a suitable method to measure local strain, a single sample can be used to correlate strain with induced effects, without the need for external sources of strain variation. These strain-induced effects have great potential for exploitation in novel devices, which may be based on bubbles themselves, or, perhaps more feasibly, on more controllable methods of straining 2D materials, such as transfer onto patterned substrates.[31]

From a different perspective, the ubiquity of bubbles means it is also important to understand any unintended effects they may have on devices, for quality control purposes. Particularly because the small sizes of modern devices are comparable to those of typical bubbles in van der Waals heterostructures. For large-scale production of graphene-based optical modulation devices to become viable, knowledge of the role that bubbles may play in device performance is vital.

In this work, we use scattering-type scanning near-field optical microscopy (sSNOM) to probe the nanoscale IR response of a network of variously shaped bubbles in an hBN-encapsulated graphene heterostructure. This reveals distinct domains with significantly enhanced absorption within individual bubbles. The boundaries of these domains correlate with ridges in the shape of the bubbles, which leads us to attribute them to nanoscale variations in strain configuration. We investigate this further by using confocal Raman spectroscopy, along with vector decomposition analysis,[32–34] to create spatial maps of strain and doping variations from the same heterostructure. We demonstrate that networks of bubbles induce mixed and intricate strain configurations, with localised areas of mostly uniaxial or biaxial strain, and that there is a pronounced increase in hole doping induced by the contaminants in the bubbles.

The encapsulated graphene heterostructure in this work was fabricated on 290 nm $SiO_2$ using mechanically exfoliated crystals of hBN and graphene. The heterostructure consists of ~220 nm-thick lower hBN, single layer graphene (SLG), and ~1.2 nm-thick upper hBN and was made using the now standard dry peel transfer technique[35] with a bespoke micromanipulation setup[36] (see **Methods**). The optical image in **Figure 1**a shows a large amount of bubbles, of varied size and shape, in the encapsulated region of the heterostructure. Zooming in on an area of the heterostructure using peak-force tapping atomic force microscopy (AFM) reveals that these bubbles are connected by thin wrinkles or filaments (**Figure 1**b). These filaments are related to the shape of the bubbles, with the number of filaments connected to each bubble corresponding to the number of sharp corners in its footprint. This implies that the proximity of the bubbles in this heterostructure may have led to their exerting a collective influence on each other's shape during formation. It is also revealed that there are variations in shape within individual bubbles. The bubbles shown have ridges in their sides, which separate regions of different curvature. There are also smaller bubbles, which are more circular in shape and are not connected by filaments.

The inset to **Figure 1**a shows a schematic cross-section of a bubble in this heterostructure, with the contamination between the lower hBN and graphene. It is also possible for bubbles to exist between the graphene and upper hBN. It would be difficult to distinguish these types of bubble using AFM alone, but we show below that the bubbles in this work strain the graphene, so we conclude that the contamination is beneath the graphene layer.

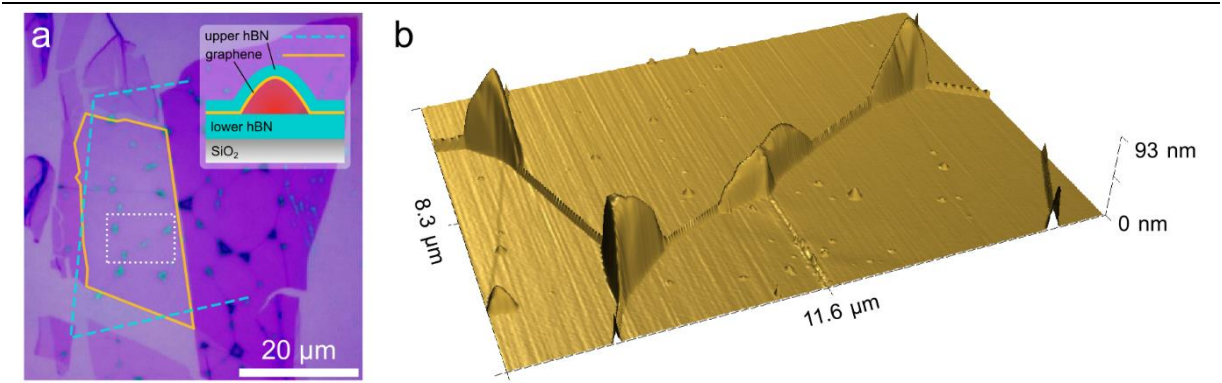

**Figure 1.** Bubbles in an hBN-graphene-hBN heterostructure. (a) An optical microscope image of the heterostructure showing bubbles between the layers. The ~220 nm lower hBN layer fills the entire view; the known edges of the single layer graphene (solid yellow line, determined by confocal Raman) and the ~1.2 nm upper hBN layer (dashed blue line, determined by AFM) are indicated. The darker purple areas are multilayer graphene. The region focused on in the rest of this work is indicated by a dashed white rectangle. Inset: side view schematic of a bubble, showing contamination (in red) trapped beneath graphene and hBN. (b) A 3D surface topography map of the area indicated in (a). The scale of the z-coordinates has been exaggerated to show topographical features more clearly.

It is important to understand the nanoscale IR properties of graphene and hBN, so they may be exploited in devices. To this end, we use sSNOM to probe how the presence of bubbles affects the IR response of the heterostructure. In this technique, IR light is focused onto a metallised AFM tip, exciting a tightly confined near-field around the tip's apex. The scattered light from the near-field interacting with the sample can then be measured. This allows imaging of complex light-matter interactions at a resolution many orders of magnitude below the diffraction limit, typically ~30 nm (see **Methods**).[37]

We show below that within graphene bubbles there exist nanoscale domains whose optical absorption is significantly altered at incident wavenumbers, $k$, of around 1000 cm$^{-1}$. These domains are not reproduced at 1362 cm$^{-1}$ which shows the effect is wavelength dependent.

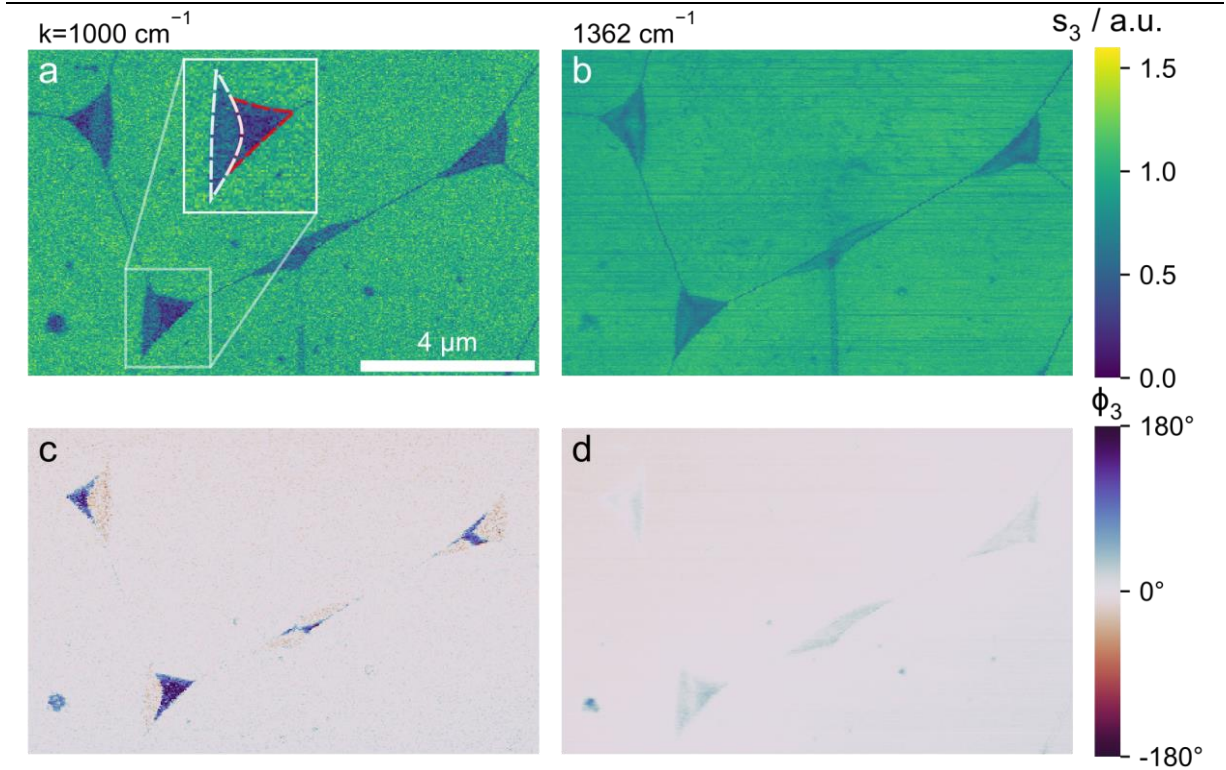

**Figure 2**. sSNOM images of nanoscale optical domains within hBN-encapsulated graphene bubbles. (a, b) Third harmonic sSNOM amplitude ($s_3$) maps, at $k$=1000 and 1362 cm$^{-1}$ respectively. Values are normalised so that the median value in each map is equal to 1. Inset of (a) shows zoom on a single bubble with domains outlined. (c, d) Third harmonic sSNOM phase ($\phi_3$) maps. Values are normalised so that the median value in each map is equal to 0°. Scale bar in (a) shared by all images.

**Figure 2** shows the third harmonic near-field scattering from the area under study (see **Figure 1**), taken at $k$=1000 and 1362 cm$^{-1}$. **Figures 2**a and **2**b show the scattering amplitude, $s_3$, normalised to the background graphene value by dividing by the median amplitude for each map, so that the background amplitude is ~1. At both 1000 and 1362 cm$^{-1}$, the bubbles and filaments have a reduced scattering amplitude, however the reduction is less significant at 1362 cm$^{-1}$. At 1000 cm$^{-1}$, there are two distinct levels of contrast observed within the larger bubbles, with one level at an amplitude of ~0.6 and the other at ~0.25 times the background amplitude (shown by red and white outlines, respectively, in the zoomed portion of **Figure 2**a). These are most visible in the two bubbles at the left of the image. They form domains whose boundaries correlate well with the topographic ridges of **Figure 1**b. The amplitude within the bubbles at 1362 cm$^{-1}$ is more homogeneous and the same pronounced domains are not observed.

**Figures 2**c and **2**d show the corresponding complex phase of the scattered light, $\phi_3$, normalised to the background graphene by subtracting the median phase for each map. At $k$=1000 cm$^{-1}$ the domains seen in **Figure 2**a are well reproduced, with multiple domains clearly visible in all large bubbles, again correlating with the geometrical ridges seen in **Figure 1**b. The parts of the bubbles with the greatest reduction in $s_3$ have a significant phase shift of ~90°, which indicates that those parts of the bubbles strongly absorb light at 1000 cm$^{-1}$. The remaining parts of the same bubbles are characterised by a slight negative phase shift of ~5°. Conversely, the phase shifts within bubbles at 1362 cm$^{-1}$ are much smaller, at ~30°, and the same pronounced domains are not observed.

To investigate the wavelength dependence in more detail, a single bubble (from the lower left corner of the image) was imaged repeatedly, while changing $k$ from 960 to 1040 cm$^{-1}$. **Figures 3**a-e show the resulting third harmonic near-field scattering amplitude maps from the bubble and the surrounding area.

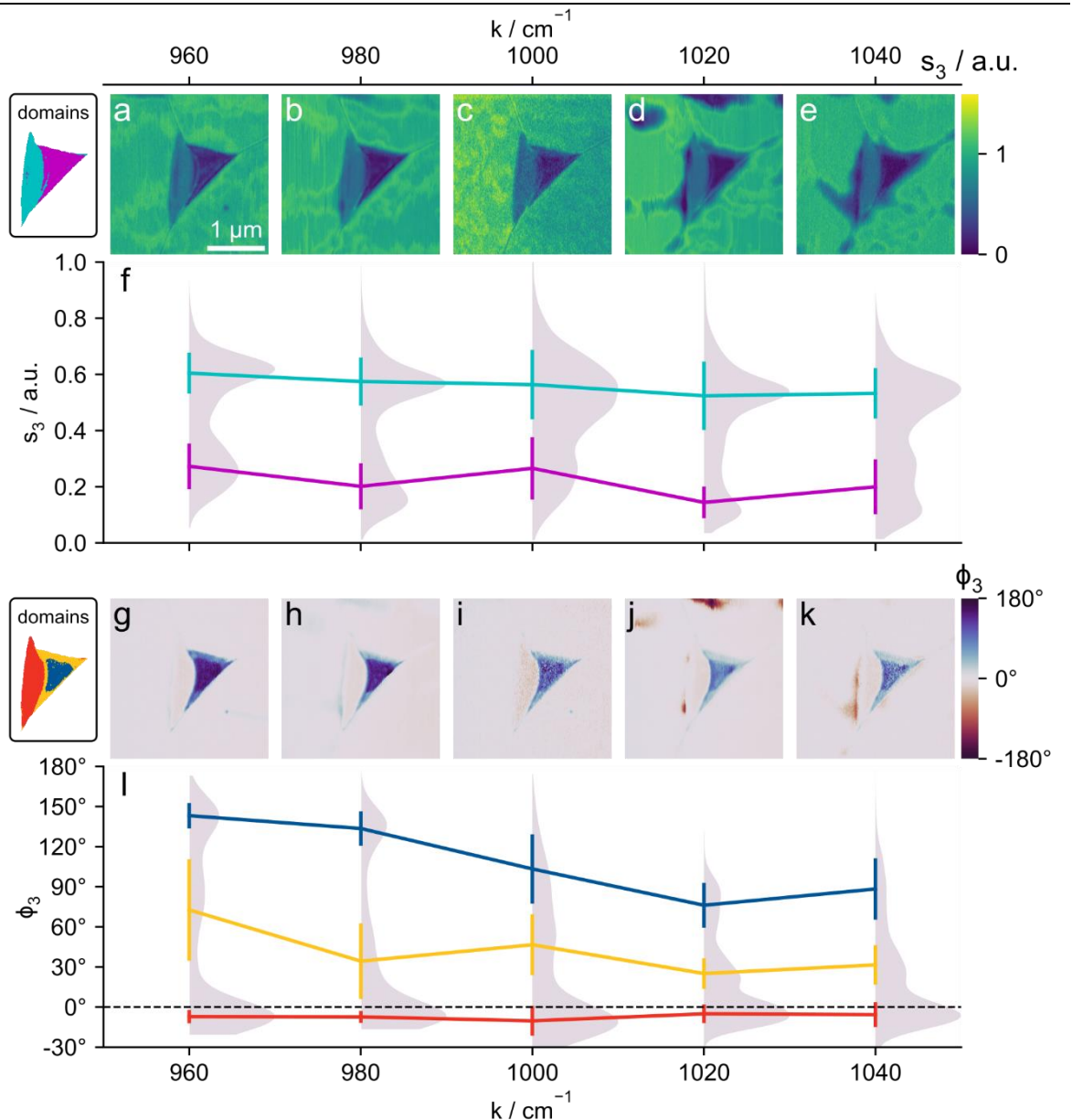

**Figure 2**. Spectroscopic dependence of domains within a bubble. (a-e) Third harmonic sSNOM amplitude ($s_3$) maps from a bubble, at wavenumbers between 960 and 1040 cm$^{-1}$. Left: Domains returned by fitting a two-component Gaussian mixture model (GMM) at 960 cm$^{-1}$. (f) Histograms showing the distribution of $s_3$ values at each wavenumber, overlaid with error bars showing the means and standard deviations returned by fitting the GMM to the distribution. Lines are coloured to match the domains shown left of (a-e). (g-k) Corresponding sSNOM phase ($\phi_3$) maps. Left: Domains returned by fitting a three-component GMM at 960 cm$^{-1}$. (l) Histograms showing the distribution of $\phi_3$ values at each wavenumber, overlaid with error bars showing the means and standard deviations returned by fitting the GMM to the distribution. Lines are coloured to match the domains shown left of (g-k). Scale bar in (a) shared by all images.

The bubble is again split into two domains. To illustrate these internal domains more clearly, and to remove the influence of surface adsorbants, which were present on the flat area, the $s_3$ values from within the bubble were isolated (see **Methods**). These values are shown for each $k$ value by histograms in **Figure 3**f.

To gain a more quantitative measure of the evolution of these domains, a Gaussian mixture model (GMM) was fit to each distribution. This assumes that a dataset is composed of $N$ normally distributed clusters, then uses the expectation-maximisation algorithm[38] to determine the parameters of the $N$ Gaussian peaks that best describe these clusters. Here $N$ was set to match the number of experimentally observed domains.

For the $s_3$ maps we observed two domains experimentally, so used $N=2$ for the GMM. The means and standard deviations returned by the GMM are overlaid as error bars in **Figure 3**f. To verify that the components returned are the same as the experimentally observed domains, and to serve as a visual key, the $s_3$ points were evaluated according to which Gaussian component they are most likely to belong to and coloured accordingly. This is displayed, using $k=960$ cm$^{-1}$ as an example, to the left of the plot.

Across the range of wavenumbers studied, the complex amplitudes of each domain do not vary significantly, with the left side remaining at ~0.6 and the right-hand side at ~0.25 times the background amplitude, as seen in **Figure 2**a.

**Figures 3**g-k show the corresponding third harmonic near-field scattering phase maps. In these images (captured at higher resolution than in **Figure 2**) there are three, rather than two, distinct levels of phase contrast within this bubble. The left side has a small negative phase shift (relative to the zero-normalised background), and the right-hand side has a high shift in the centre and reduced shift at the edges. The phase shifts in the right-hand side reduce with increasing $k$.

The same process of isolating values from the bubble, plotting the distribution and fitting to the distribution with a GMM was repeated for the phase images, this time with $N=3$. The resulting data is plotted in **Figure 3**l. The domain with a slight negative phase shift of around 5° is unchanged by $k$. However the domains with high phase shift are wavenumber dependent, with the centre right domain shifting from ~150° to ~90° and the edges of the right-hand side shifting from ~75° to ~30°, with a $k$ increase from 960 to 1040 cm$^{-1}$. This may indicate that this range of wavenumbers is on the side of an absorption peak for this area of the bubble.

These domains are separated by ridges in the bubbles' shape, and the shapes of bubbles are known to result from competition between van der Waals and elastic potential energies. For this reason it is probable that the domains will have different strain configurations. To investigate this further, we used Raman spectroscopy to visualise the strain variations in this heterostructure.

Raman is routinely used as an indicator of the quality of graphene.[39] The solid blue line in **Figure 4**a shows a Raman spectrum taken from a flat area of the heterostructure, free from bubbles. It displays the characteristic G and 2D peaks of graphene, at ~1580 and ~2680 cm$^{-1}$ respectively. The 2D to G height ratio of ~3 and 2D peak full width at half maximum (FWHM) of ~20 cm$^{-1}$ are characteristic of high-quality hBN-encapsulated SLG. The graphene D peak at ~1362 cm$^{-1}$ is not apparent, which is another indicator of defect-free graphene.

Analysis of these Raman peaks can provide a wealth of information about the strain and doping of graphene. To produce spatial maps of these quantities, a Raman datacube was collected from the area of the sample shown above, and Lorentzians were fit to the G and 2D peaks. The extracted positions for the peak centres, $\omega_G$ and $\omega_{2D}$, are correlated in a scatterplot in **Figure 4**b.

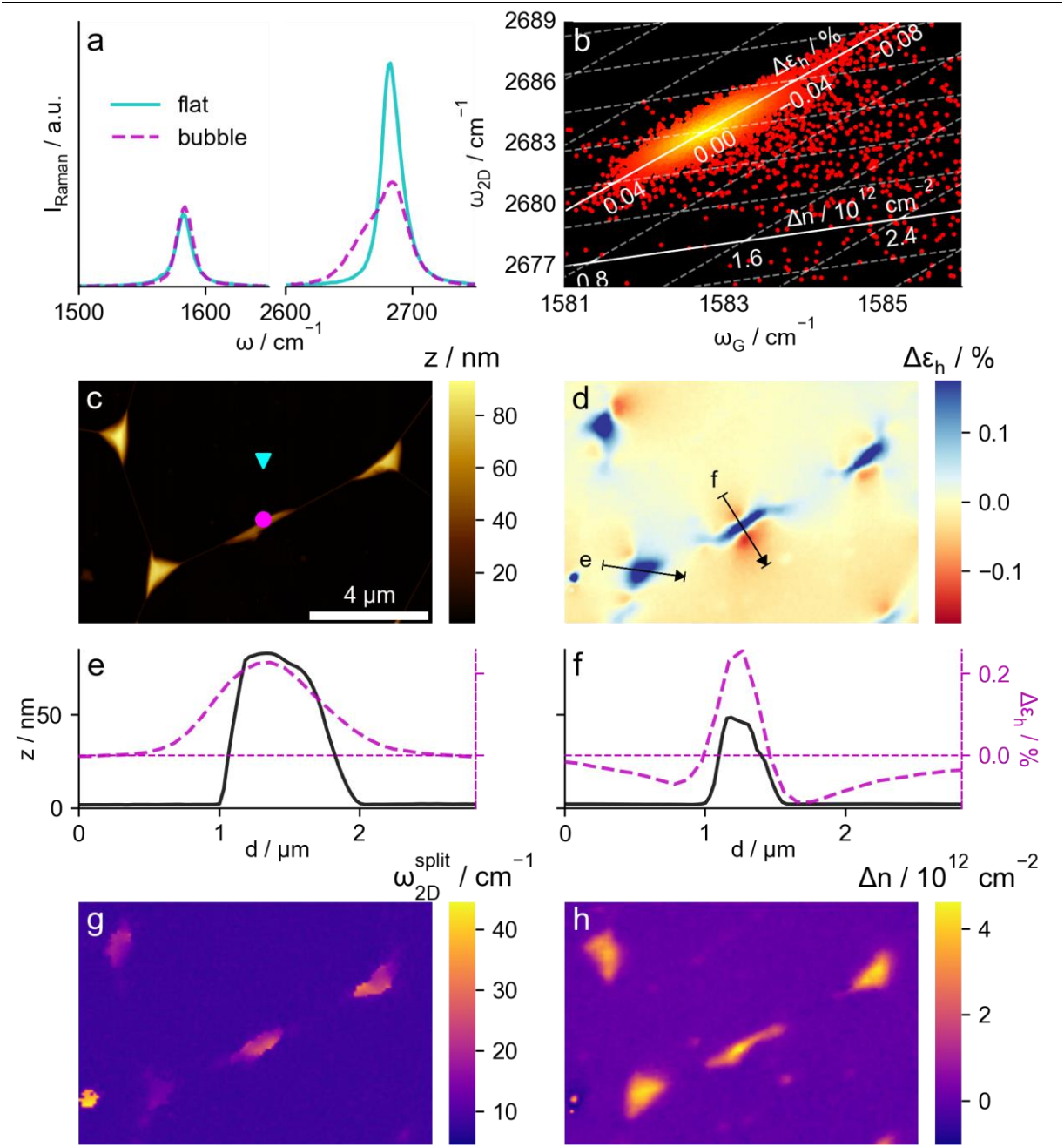

**Figure 4**. Graphene strain and doping analysis using vector decomposition model. (a) Raman spectra from a flat area (solid blue, location: triangle in (c)) and a bubble (dashed purple, location: circle in (c)) in the hBN-encapsulated graphene heterostructure. (b) G and 2D position scatterplot from the region under study, showing the model used to separate the effects of hydrostatic strain ($\varepsilon_h$) and hole concentration ($n$). The colour of the points indicates the local density in the scatterplot, where bright yellow corresponds to a higher density of points. (c) AFM topography map. (d) Median normalised map of $\varepsilon_h$ changes. (e, f) Line profiles showing height and strain across a more biaxially and uniaxially strained bubble, respectively (locations shown in (d)). (g) Map of 2D splitting, which correlates with shear strain. (h) Median normalised map of doping changes. Scale bar in (c) shared by (d), (g) and (h).

Both hole doping, $n$, and strain, $\varepsilon$, in graphene cause a shift of the G and 2D peaks, so additional analysis is needed to determine the separate $\varepsilon$ and $n$ contributions. This can be achieved by correlating the shift of both peaks using vector decomposition, and comparing them to empirical measurements.[33,40] The result of this is that the evolution of a point in an $\omega_G$-$\omega_{2D}$ correlation plot under changing $\varepsilon$ or $n$ can be approximated by a straight line with a known gradient.[32] This vector decomposition analysis is illustrated by the additional axes shown in **Figure 4**b.

Different species of strain, for example biaxial and uniaxial, result in different gradients for the straight line associated with strain changes. In samples with unknown or mixed strains this can be accounted for by choosing to use the gradient associated with the hydrostatic strain, $\varepsilon_h$. This is a component of the full biaxial strain tensor, $\varepsilon_{bi}$, which can be described by two components: $\varepsilon_h$ and shear strain, $\varepsilon_s$.[33]

$$\varepsilon_{bi} = \begin{pmatrix} \varepsilon_{xx} & \varepsilon_{xy} \\ \varepsilon_{yx} & \varepsilon_{yy} \end{pmatrix} \tag{1}$$

$$\varepsilon_h = \varepsilon_{xx} + \varepsilon_{yy} \tag{2}$$

$$\varepsilon_s = \sqrt{\left(\varepsilon_{xx} + \varepsilon_{yy}\right)^2 + 4\varepsilon_{xy}^2} \tag{3}$$

(Assuming $\varepsilon_{xy} = \varepsilon_{yx}$)

A qualitative explanation of these relationships is that $\varepsilon_h$ corresponds to an isotropic change in size of the unit cell, while $\varepsilon_s$ corresponds to a change in the shape of the unit cell, which leaves its area unchanged.

For ease of comparison with the colocalised Raman-acquired maps, the AFM topography image from **Figure 1**b is shown again as a 2D image in **Figure 4**c.

**Figure 4**d shows the median-normalised hydrostatic strain distribution around the same area of bubbles. The greatest values of tensile (positive) strain are localised at the centres of bubbles, coinciding with the areas of greatest height. This is consistent with expectations for bubbles with contamination beneath the graphene layer. Interestingly the areas of graphene and hBN in the vicinity of the bubbles are not free from strain. The areas between bubbles, close to the filaments (**Figure 1**b), show small increases in tensile strain relative to the background.

Additionally, at the sides of some bubbles, particularly the narrow, elongated bubble in the centre of the image, there are small areas of more compressive strain. These are due to Poisson contraction, a phenomenon associated with uniaxial strain configurations.[41] This is illustrated more clearly by the AFM and $\varepsilon_h$ line profiles taken from a more biaxially (**Figure 4**e) and a more uniaxially (**Figure 4**f) strained bubble, as indicated in **Figure 4**d.

Uniaxial strains cause an anisotropic deformation to the lattice, and therefore correlate with $\varepsilon_s$. This anisotropy causes a polarisation dependent splitting of the peaks about their centres.[33] The dashed purple line in **Figure 4**a shows a Raman spectrum taken from the central bubble in the heterostructure (indicated in **Figure 4**c). The presence of $\varepsilon_s$ has caused splitting of the 2D peak. To minimise the polarisation dependence, we used circularly polarised light for the incident Raman laser. However there is a small residual polarisation dependence, introduced by the diffraction grating, which explains the asymmetry of the split 2D peak.

By fitting two Lorentzians to a peak, it is possible to obtain a measure of this peak splitting. We performed this for the 2D peak, and the resulting splitting, $\omega_{2D}^{split}$, is shown in **Figure 4**g. As discussed above, this should be proportional to $\varepsilon_s$. Indeed the greatest values of $\omega_{2D}^{split}$ correlate well with the areas of Poisson contraction in **Figure 4**d, providing further evidence that these areas have a more uniaxial strain configuration.

The change in hole concentration of the graphene is also returned by the vector analysis. **Figure 4**h shows the median-normalised doping distribution. The concentration correlates well with the topography of the bubbles, showing an increase of ~$4\times10^{12}$ cm$^{-2}$ relative to the background at the bubble locations. The influence of bubbles on the carrier concentration is much more localised than on $\varepsilon_h$. Doping is seen even from the small, more rounded bubbles, not connected by filaments.

Compared to the strain distribution, the doping is more tightly confined to the bubble locations. This confirms that the primary doping source in these bubbles is the contaminant that fills them, and that the doping and strain here are independent. The graphene between the bubbles is shown to be dopant-free, which supports the perception that bubble formation is an effective self-cleaning mechanism.[29,30]

We do not observe features that correlate with the domains seen from sSNOM in any of the Raman-acquired maps. However this is to be expected, as the spot size for our Raman measurements (~450 nm) is comparable to the lateral sizes of the domains (~500 nm).

The vector decomposition model relies on a few assumptions about the nature of the graphene, namely that it is single layer, relatively defect free and primarily p-type doped.[32] The validity of the first two assumptions for our heterostructure is verified by Raman spectroscopy, as shown above, but the type of doping (n- or p-) cannot be determined from the peak shift alone. However bubble-free graphene encapsulated in hBN is known have an intrinsic doping close to charge neutrality, due to a lack of dangling bonds in the hBN and screening of the graphene from charged impurities and atmospheric dopants.[42–45] Typical hydrocarbon contaminants in bubbles induce hole doping in graphene,[29,30,43] so we assume that the primary doping mechanism is p-type.

The strain and doping maps presented above are both normalised so that their median value is equal to 0. This is to compensate for dielectric screening of the graphene Kohn anomaly, caused by hBN, which adds a constant offset to both $\omega_G$ and $\omega_{2D}$.[46,47] This results in expected behaviour for encapsulated graphene in the $\Delta n$ map, with flat areas close to charge neutrality and increased p-type doping at bubble locations, as discussed above, so we may make the approximation that $\Delta n \approx n$.

It is known that variations in in Fermi level, $E_F$, can change the IR absorption of graphene. We used the above approximation to calculate the Fermi level, according to the following equation. [48]

$$E_F = \frac{h}{2\pi} v_F \sqrt{\pi n} \qquad (4)$$

Here $h$ is Planck's constant, and $v_F$ (≈$10^6$ m s$^{-1}$) is the graphene Fermi velocity. Applying this to the Raman-acquired $n$ values yields a map of $E_F$, shown in **Figure 5**a.

The absorption of graphene is proportional to the real part of its frequency dependent optical conductivity, $\sigma(\omega)$. At high photon energy, $E_{ph}$, interband transitions dominate, leading to a universal conductance value of $\sigma_0 = \pi e^2/2h$, where $e$ is the elementary charge. This results in a flat absorption of ~2.3%. But for photon energies below $2E_F$, these interband transitions are prevented by Pauli blocking. This leads to a drop in the absorption and a Drude-type response to light. As $E_F \propto \sqrt{n}$, the position of the Pauli blocking transition, at $E_{ph}=2E_F$, is dependent on the level to which the graphene is doped.[4]

The optical conductivity of graphene as a function of $E_F$, calculated using the local random phase approximation[4] (see **Methods**) at a temperature, $T=0$ K, is shown for wavenumbers of 1000 and 1362 cm$^{-1}$ in **Figure 5**b. The onset of the Pauli blocked regime occurs at $E_F=62$ and 84 meV respectively. To more easily compare this calculation with the experimentally determined $E_F$ values, the values from **Figure 5**a are shown as a histogram in **Figure 5**b.

At both wavenumbers, $E_F$ is below the Pauli transition for the areas of flat graphene (corresponding to the prominent peak in the histogram), which indicates that they should be in the high absorption, universal conductance regime. The doping at the bubbles shifts $E_F$ into the Pauli blocked regime, which should be accompanied by a reduced $\sigma$ and absorption, as well as difference in $\sigma$ between 1000 and 1362 cm$^{-1}$. This may explain why the strongly absorbing domains appear only at the lower wavenumber. However, the fact that we see an increase, rather than the predicted decrease, in absorption at the bubble locations tells us that we cannot explain all the variations from $E_F$ alone, and that other factors need to be considered.

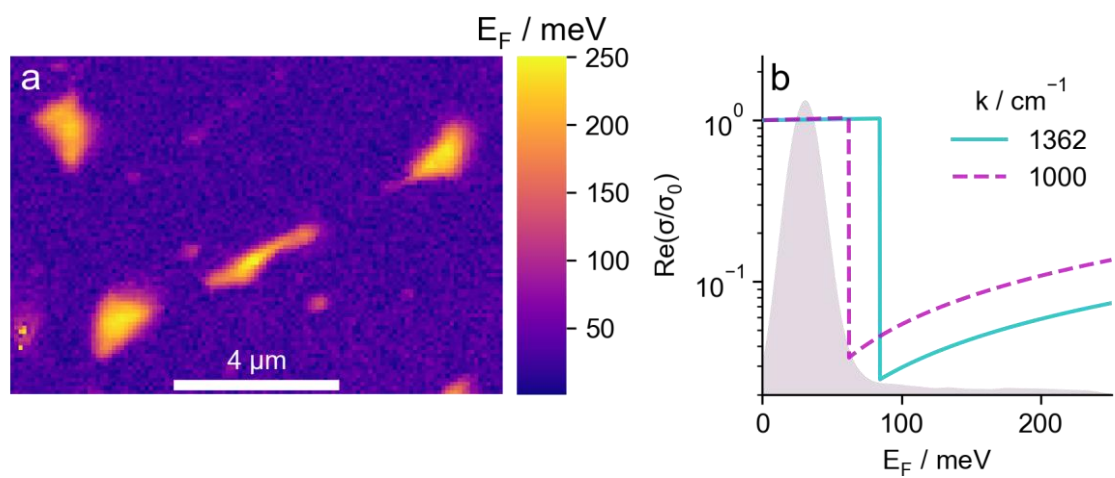

**Figure 5**. Graphene Fermi level and onset of Pauli blocking. (a) Raman-acquired map of $E_F$, calculated from the data in **Figure 4**h. (b) Real part of graphene optical conductivity, normalised to the universal conductance value, $\sigma_0$. Calculated using the local random phase approximation at $T=0$ K.[4] Background: normalised kernel density histogram of $E_F$ values in (a) (linear y-scale).

Graphene-hBN heterostructures, doped above the Pauli transition are known to support SPPs in the wavenumber region around 1000 cm$^{-1}$.[1,2,4,16,17] This is the case for the bubbles in this heterostructure. We do not observe the plasmonic standing waves typical of sSNOM measurements of SPPs in graphene,[3–10] however this may be due to the lack of low-loss reflection sites for SPPs, such as graphene edges. Nevertheless, it is possible that the origin of the domains we observe at 1000 cm$^{-1}$ could be plasmonic in nature.

Theoretical simulations of self-assembled 3D graphene nanostructures, such as pyramids and polyhedrons, have suggested that vertexes and edges in such structures could have strong plasmonic field enhancement due to combined in- and out-of-plane coupling of plasmonic modes [49,50]. We do not see evidence of this enhancement at the edges and vertexes in the bubbles in this work. This may be because the bubbles we studied (which have heights ~0.1 × their footprints) are much flatter than the 3D self-assembled structures (which have comparable heights to their footprints). This would reduce the effect of the mode coupling.

Nanoscale strain differences induced in the graphene by bubbles are a likely candidate for the origin of the optical domains. Though the Raman spot size prevents us from visualising strain variations within individual bubbles, we have shown that, at a larger scale, the strain configuration from a network of closely spaced bubbles is intricate and varied, and not simply related to height. The distributions we have measured, featuring Poisson contraction and increased tensile strain between bubbles, are similar to theoretical calculations of stress in graphene sheets deposited over networks of trapped particles.[51] Knowing that strain minimisation is one of the main factors governing a bubble's shape,[25,51] the topographic ridges separating the optical domains lead us to conclude that the different domains must have different strain configurations.

The bandstructure of graphene is significantly altered by strain,[22,52–56] which may explain the different absorption in different domains. If the origin of the domains is indeed plasmonic, then they may be caused by an alteration of the plasmonic dispersion brought about by bandstructure changes between differently strained areas.[17,27,57–60]

Strain-induced areas of enhanced light absorption have great potential for the design of future graphene-based IR devices. Networks of bubbles like the one shown here are generally created stochastically during material transfer and are therefore difficult to control. But it may be possible to create similar strain patterns in more reproducible ways, for example by inflating graphene bubbles by pumping pressurised gas through networks of holes in a substrate,[61] or by depositing graphene onto an array of nanopillars.[31,62] In the latter example, one could imagine creating a photonic crystal, or metamaterial, with spatially modulated absorption, simply by depositing graphene onto a prepatterned substrate.

In conclusion, we have shown that within bubbles in closely spaced networks in graphene-hBN heterostructures there exist nanoscale domains with strongly enhanced absorption of IR light at 1000 cm$^{-1}$. We demonstrated with Raman spectroscopy and vector decomposition analysis that these networks induce intricate and varied strain configurations in graphene. Furthermore, we deduced that, due to shape variations seen in AFM, the IR domains must also have nanoscale differences in strain configuration. This led us to attribute the origin of the domains to these strain differences, which could alter the plasmonic dispersion of the graphene. This result will have profound consequences for the design and quality control of future graphene-based IR devices.

**Methods.** *Sample fabrication.* Graphite (obtained from HQ Graphene) was mechanically exfoliated by peeling with Nitto Denko BT-150E-CM tape before being pressed onto a Si/SiO$_2$-290 nm wafer (heated to 60°C to improve adhesion). Suitable monolayer graphene was then identified using optical microscopy and picked up with few-layer hBN using the poly(methyl-methacrylate) (PMMA) dry peel transfer technique[35] using a bespoke micromanipulation setup.[36] Following this, the sample on the membrane was stamped onto a bulk (> 30 nm) crystal of laterally large (> 50 × 50 μm) hBN which had previously been exfoliated onto a Si/SiO$_2$-290 nm wafer. The completed structure therefore had the form hBN/Gr/hBN/SiO$_2$/Si.

After full encapsulation, the sample was annealed at 300°C for 3 hours in a high vacuum environment (> $10^{-7}$ mbar) to remove polymer residue on the surface of the heterostructure and promote the self-cleaning mechanism[29] and subsequent bubble formation.[25]

*Atomic Force Microscopy.* The AFM image shown was taken on a Bruker Dimension Icon SPM, using silicon nitride AFM tips with a resonant frequency of ~300 kHz. The heterostructure was imaged using peak-force tapping mode, with a peak force setpoint of 500 pN to ensure that the tapping was gentle and didn't disturb the bubbles.

*sSNOM.* The sSNOM measurements were performed on an Anasys Instruments NanoIR-2s AFM, with excitation provided by a quantum cascade laser (QCL). PtIr coated AFM tips with a resonant frequency of ~285 kHz were used. IR light from the QCL was focused onto the AFM tip using a parabolic mirror, exciting a tightly confined near-field around the tip apex. The same mirror was used to collect the light scattered from the tip and sample, which was measured with a mercury cadmium telluride detector. To isolate the component of the total scattered light which is caused by the near-field interacting with the sample, the AFM was operated in tapping mode and a lock-in amplifier was used to demodulate the detected light at the third harmonic of the tapping frequency.[37]

To capture the full complex information, the light was collected via a Michelson interferometer. The sample was imaged twice, with the interferometer reference mirror fixed at two orthogonal phases, and the images were combined to yield the complex amplitude and phase. For each image, an additional pass was taken with the interferometer reference arm blocked, which was then subtracted from the image to account for any self-homodyne background.[37,63]

To combine consecutive images, any lateral thermal drift of the sample position was determined using cross correlation of the topography images, then the images were translated and cropped to show the same area of the sample (to the nearest pixel).

To isolate the $s_3$ and $\phi_3$ values of the bubble from the flat areas in **Figure 3**, we used only values whose corresponding height, determined by AFM, was greater than 15 nm. We attribute the variations in $s_3$ and $\phi_3$, seen on the flat areas, to loose material on the sample surface, which we could see in the AFM topography. We did not observe this contamination on the bubbles themselves (likely due to the steeper sides), meaning the influence of this material was effectively removed by selecting only values from within the bubble.

To verify that the observed domains were not the result of tip shadowing, or other direction sensitive effects, bubbles were imaged in multiple orientations. The same domains were observed consistently.

*Raman Spectroscopy.* Raman mapping was performed using a Renishaw inVia confocal Raman microscope. We used a 532 nm excitation laser focused through a 100× objective, with a numerical aperture of 0.85. The laser power incident on the sample was ~0.5 mW. A 1800 line mm$^{-1}$ diffraction grating was used. The measurements were taken in high confocality mode, resulting in an estimated spot size of ~450 nm in FWHM.

A quarter-wave plate was used to create circularly polarised light in the incident beam to reduce the polarisation dependency of the measurements and therefore better resolve the splitting of the 2D peak.[33] There is still a small residual polarisation dependence from the spectrometer, typically less than ~10%.

For extracting strain and doping information from the shifts of the G and 2D peaks, single Lorentzians were fit to each peak to extract the positions. The gradients used to distinguish between carrier concentration and hydrostatic strain, as well as the Grueneisen parameter used to convert peak shift in cm$^{-1}$ to percentage strain were taken from a 2017 work by Mueller *et al.*[33] The empirical measurements used to convert peak shift in cm$^{-1}$ to hole concentration in cm$^{-2}$ were taken from a 2015 work by Froehlicher and Berciaud.[40]

*Optical Conductivity Calculations.* The optical conductivity of graphene was calculated using the local random phase approximation in the limit where $T$=0 K, as described in the Supporting Information of ref. 4. We used an estimated carrier scattering time of $\tau$=10$^{-13}$ s, taken from the same work.

# AUTHOR INFORMATION
**Corresponding Author**
*Email: olga.kazakova@npl.co.uk
**ORCID**
Tom Vincent: 0000-0001-5974-9137
Matthew Hamer: 0000-0003-3121-6536
Irina Grigorieva: 0000-0001-5991-7778
Vladimir Antonov: 0000-0002-0415-5267
Alexander Tzalenchuk: 0000-0001-7199-9331
Olga Kazakova: 0000-0002-8473-2414
**Notes**
The authors declare no competing financial interest

# ACKNOWLEDGEMENTS
This project has received funding from the European Union's Horizon 2020 research and innovation programme under grant agreement GrapheneCore2 785219 number. The work has also been financially supported by the Department for Business, Energy and Industrial Strategy though NMS funding (2D Materials Cross-team project).

TABLE OF CONTENTS ONLY

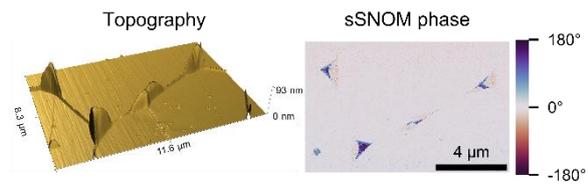